\documentclass{aa}  
\usepackage{graphicx}
\usepackage{txfonts}
\usepackage{natbib}
\bibpunct{(}{)}{;}{a}{}{,}
\usepackage{amsmath}
\usepackage{array}
\usepackage{float}
\usepackage{import}
\usepackage{bibunits}
\usepackage[version=4]{mhchem}
\usepackage{siunitx}

\begin{document}

\title{Formation of methane and cyclohexane through the hydrogenation of toluene}
\titlerunning{Formation of \ce{CH4} and \ce{C6H12} through the hydrogenation of toluene}

   \author{A. T. Hopkinson
          \inst{1}\fnmsep
          \and
          F. G. Doktor 
          \inst{1}\
          \and
          J. Pitfield 
          \inst{1}\
          \and
          M. Moll 
          \inst{3}\
          \and
          J. D. Thrower 
          \inst{1}\
          \and
          L. Hornek{\ae}r 
          \inst{1,2}\
          }

   \institute{Center for Interstellar Catalysis, Physics and astronomy Department, Aarhus University, Aarhus, 8000, Denmark\\
              \email{a.hopkinson@phys.au.dk}
              \email{thrower@phys.au.dk}
         \and
             Interdisciplinary Nanoscience Center, Aarhus University, Aarhus C, 8000, Denmark
         \and
             Heidelberg University, Department for Physics and Astronomy, Heidelberg, 69117, Germany\\
             }

   \date{}

  \abstract
   {}
   {Methylated polycyclic aromatic hydrocarbons (PAHs) have been hypothesised to be present in the interstellar medium (ISM) through their 3.4 and 6.9\,$\mu$m absorption bands. To investigate the hydrogenation of these methylated PAHs, toluene (\ce{CH3C6H5}), as a simple analogue of such species, was exposed to H-atoms. This demonstrated how the presence of a methyl group changes the reactivity towards atomic hydrogen as compared to benzene and larger aromatic species such as PAHs and how this may alter its chemistry in the ISM.}
   {Toluene was deposited onto a graphite surface in an ultrahigh vacuum (UHV) chamber and then exposed to a H-atom beam. Temperature programmed desorption (TPD) measurements were used to investigate the reaction between H-atoms and toluene. The masses of hydrogenation products were measured with a quadrupole mass spectrometer (QMS).}
   {H-atom exposure of toluene leads to superhydrogenation of toluene and the formation of methyl-cyclohexane (\ce{CH3C6H11}) at long exposure times. The initial cross-section of H-addition is smaller than that for larger PAH molecules. Methyl-cyclohexane can be further hydrogenated, leading to the detachment of the methyl group and production of cyclohexane (\ce{C6H12}) and methane (\ce{CH4}).}
   {Toluene may be fully hydrogenated through its interaction with H-atoms, although it has a smaller initial cross-section for H-atom addition compared to larger PAHs. This likely reflects it having a smaller geometric cross-section and the low flexibility of the benzene ring when undergoing sp$^3$ hybridization. The removal of the methyl group at high H-atom fluences provides a top down formation route to smaller molecules with the possibility of the formation of a radical cyclohexane combining with other species in an interstellar environment to form prebiotic molecules.}

   \keywords{astrochemistry – molecular processes – methods: laboratory: molecular – methods: laboratory: solid state – ISM: molecules
               }

\maketitle
   
\section{Introduction}

5-10 percent of carbon in the interstellar medium (ISM) is expected to be locked up in polycyclic aromatic hydrocarbons (PAHs) \citep{RevModPhys.85.1021}. The aromatic infrared Bands (AIB) observed in the diffuse ISM at 3.3, 6.7, 7.7, 8.6 and 11.3\,$\mu$m, corresponding to predominately aromatic species \citep{Tielens2008}, have been attributed to the presence of interstellar gas-phase PAHs. Recent progress has been made in the identification of specific PAHs via their cyano derivatives, making radio observations possible due to the presence of a strong dipole moment. These observations have revealed the presence of a range of simple PAHs and related species such as benzonitrile (\ce{C6H5CN}) \citep{cyclobenzenebret,Burkhardt2021a} as well as the four ringed, nitrogen substituted 1-cyanopyrene \citep{Wenzel2024a} and 2-cyanopyrene\citep{Wenzel2024b} and the seven-ring cyanocoronene (\ce{C24H11CN}) \citep{Wenzel_2025}. The detection of such substituted PAHs strongly suggests the presence of their pure hydrocarbon equivalents, consistent with the detection of indene (\ce{C9H8}) in TMC-1 \citep{burkhardt2021b}.

PAHs are, in general, expected to play an important role in contributing to complex interstellar chemistry through top down processes, such as their fragmentation via energetic processing to form \ce{H2}, \ce{C2H2} and \ce{C2H4} \citep{marciniak2021photodissociation,2022A&A...663A.150T,simonsen2024molecular}. PAHs have been shown experimentally to interact with atomic hydrogen to form superhydrogenated PAHs (HPAHs), which can act as catalysts for \ce{H2} formation through abstraction reactions \citep{C3CP54073A,C3FD00151B,Skov2016TheIO,magicnumbers,2019MNRAS.486.5492J}. This process has been suggested to contribute to H$_2$ formation in photodissociation regions (PDRs), where a higher rate of H$_2$ formation is required to account for the observed H$_2$ abundance \citep{habart2004some}. This is supported by an observed correlation between \ce{H2} and PAH emission in the $\rho$ Oph-W PDR \citep{Habart2003}. Recently, James Webb Space Telescope (JWST) observations of the Orion Bar PDR have revealed how PAHs exist in locally deuterated or superhydrogenated states indicating PAH interactions with the atomic gas \citep{peeters2024pdrs4all}. In addition to the strong aromatic features observed in the IR observations, aliphatic vibrations at 3.4 and 6.9\,$\mu$m have largely been attributed to the presence of HPAHs where the aromaticity has been broken through hydrogen addition, leading to the presence of aliphatic, sp$^3$ hybridized, groups. It has also been suggested that these bands could result from methylated PAHs \citep{Sandford_2013, Steglich_2013}. Methyl PAHs have also been suggested to play a role in the growth of PAH molecules through the methyl addition/cyclization (MAC) mechanism \citep{Shukla2010} in which cycles of \ce{^{.}CH3} and H-atom abstraction lead to ring growth.  Similarly, the methylidyne addition–cyclization–aromatisation (MACA) mechanism has been proposed as a low temperature route to PAH molecules containing five membered rings such as corannulene, proceeding through \ce{^{.}CH} addition \citep{Doddipatla2021}.

Within the solar system, toluene has been identified on the comet 67P/Churyumov-Gerasimenko with the use of \textit{in situ} mass spectrometry \citep{toluene67p} and it has also been identified in the atmosphere of Saturn's moon Titan by the ion and neutral mass spectrometer on the Cassini spacecraft \citep{MAGEE20091895}. It has been suggested, on the basis of UV emission and mid-infrared spectroscopy, that toluene might be present in comets, both from the Jupiter family and the Oort cloud \citep{2023JApA...44...89V}, although this has recently been challenged by \citet{Rouille2025}. Furthermore, it has been shown experimentally and through theoretical calculations that toluene can form benzonitrile, tolunitriles and polycyclic aromatic nitrogen containing hydrocarbons (PANHs) when mixed with cyano radicals in ISM conditions, such as those found in the Taurus molecular cloud \citep{2023ApJ...950...55W,doi:10.1021/acs.jpca.0c06900}. Nitrogen substituted molecules such as purines which are important due to their prebiotic nature as the building blocks of DNA bases. It is therefore relevant to determine how simpler related species such as toluene react with hydrogen under ISM conditions.

Our previous studies have focused on the superhydrogenation of simple PAH molecules such as coronene with a focus on how the PAH structure determines its reactivity towards atomic hydrogen. Pentacene was used to determine how the presence of zigzag edges affects the hydrogenation process \citep{pentacine} while the curved PAH corannulene \citep{rijuthacornaulene} was used to determine the effect of the presence of a 5-membered ring. Oxygen-functionalised PAHs represent another class of substituted aromatic molecules that have been investigated to determine the impact of the oxygen group on reactivity towards hydrogen via the investigation of pentacenequinone \citep{pentquinine}. The simplest aromatic species, benzene has been shown to react similarly to PAHs with H-atom addition leading to cyclohexane \citep{benzenecopper,benzeneplat}. Here, we consider how the presence of a single methyl functional group affects the reactivity of PAHs towards H-atoms, using toluene as a simple model of a single aromatic ring with a methyl group (\ce{CH3}) attached to it. We have exposed a monolayer of toluene adsorbed on a graphite substrate to a beam of H-atoms and determined the addition cross-section and reaction products using quadrupole mass spectrometer during a subsequent thermal desorption measurement.

\section{Methods}

\subsection{Experimental}

The experimental setup and temperature programmed desorption (TPD) procedure have been described previously in detail \citep{C3CP54073A,C3FD00151B,Skov2016TheIO,2019MNRAS.486.5492J}. The experiments were conducted in an ultrahigh vacuum (UHV) chamber with a base pressure below $5\times10^{-10}$\,mbar. The substrate was a highly oriented pyrolytic graphite (HOPG) sample (SPI Grade 2) that was cleaved \textit{ex situ} prior to mounting.  HOPG was used as an interstellar carbonaceous dust grain analogue surface. The substrate is mounted onto a tantalum sample holder that can be cooled to 110\,K with liquid nitrogen. The sample temperature was measured using a C-type thermocouple junction mounted between the sample holder and the front face of the HOPG. Temperature calibration was performed at temperatures $>200^{\circ}\mathrm{C}$ with the aid of a pyrometer. This was then matched to the lower temperature region according to \citet{1996JVSTA..14..260S}. The HOPG was annealed to 1100\,K by electron bombardment on the rear side of the sample holder for an initial cleaning prior to the experiments and to 600 K before each measurement to remove any adsorbed species from the surface. 

Toluene (\ce{C6H5CH3}; TCI purity >\,$99.5\%$) was purified through multiple freeze-pump-thaw cycles under high vacuum. Toluene was deposited onto the HOPG via a fine leak valve and directed through a stainless steel tube. For consistency, and to ensure a complete exposure to H-atoms, a single monolayer of toluene was used for all measurements. The procedure for preparing a monolayer is described in Appendix \ref{tol_mono}. 

The as-deposited toluene was then exposed to a beam of H-atoms that were produced in a H-atom beam source (HABS; MBE Komponenten GmbH) \citep{HABS}. This source produces a beam of neutral H-atoms by passing \ce{H2} along a tungsten capillary heated to approximately 2000\,K. The beam was subsequently cooled to a temperature of 300\,-\,1500\,K by collisions within a custom curved quartz nozzle \citep{freddythesis}. The upper limit of this temperature range is based on the simple inelastic collision simulations, utilizing the hard cube scattering model, which does not include effects such as trapping-desorption. The atomic flux at the sample was calibrated by measuring hydrogen uptake curves on a silicon surface \citep{freddythesis} and found to be:

\begin{equation}\label{flux}
      F_\mathrm{H} = (1.2\pm0.6)\times10^{15}\text{\,cm}^{-2}\text{s}^{-1}
\end{equation}

After hydrogen exposure the reaction products were determined through a TPD measurement in which the sample was heated at a rate of 1\,K/s. The heating ramp was controlled by a cryogenic PID controller (Lakeshore Model 336). Desorbing molecules were detected with a quadrupole mass spectrometer (QMS; Extrel CMS LLC) with a mass range of 1\,-\,500 amu. Toluene has a mass of 92\,amu and so, in this experiment, a mass range of 30\,-\,180\,amu was measured to cover a range of expected product masses whilst maintaining a sufficient temperature resolution. 

\subsection{Theoretical calculations}

All theoretical calculations were performed with density functional theory (DFT) as implemented in the ORCA software package~\cite{ORCA,ORCA5}. An empirically weighted hybrid functional was used to model the electron exchange correlation interaction, consisting of a mixture of Becke's~\cite{becke1993density} three-parameter exchange functional with the Lee-Yang-Parr correlation functional~\cite{lee1988development} (B3LYP). For representation of the electron charge density, a def2-TZVP~\cite{weigend2005balanced} basis set and def2/J~\cite{weigend2006accurate} auxiliary basis set were constructed. Van der Waals effects are taken into account according to the Grimme D4 method~\cite{GrimmeD4,grimme2011effect,grimme2010consistent}. 
For each proposed reaction, an interpolated initial pathway is created according to the IDPP method outlined by~\cite{IDPP}. Start and endpoints of this path were relaxed directly according to a force tolerance of 0.025 eV\slash\r{A}. Nudged elastic band calculations with the climbing image convention were then performed (\cite{NEB0, NEB1, NEB2}) to identify the saddle point in the minimum energy pathway and corresponding barrier height of these reactions. 

\section{Results}
\subsection{Hydrogenation of the toluene monolayer}\label{tol_Haddition}

\begin{figure}[h!]
    \resizebox{0.85\hsize}{!}{\includegraphics{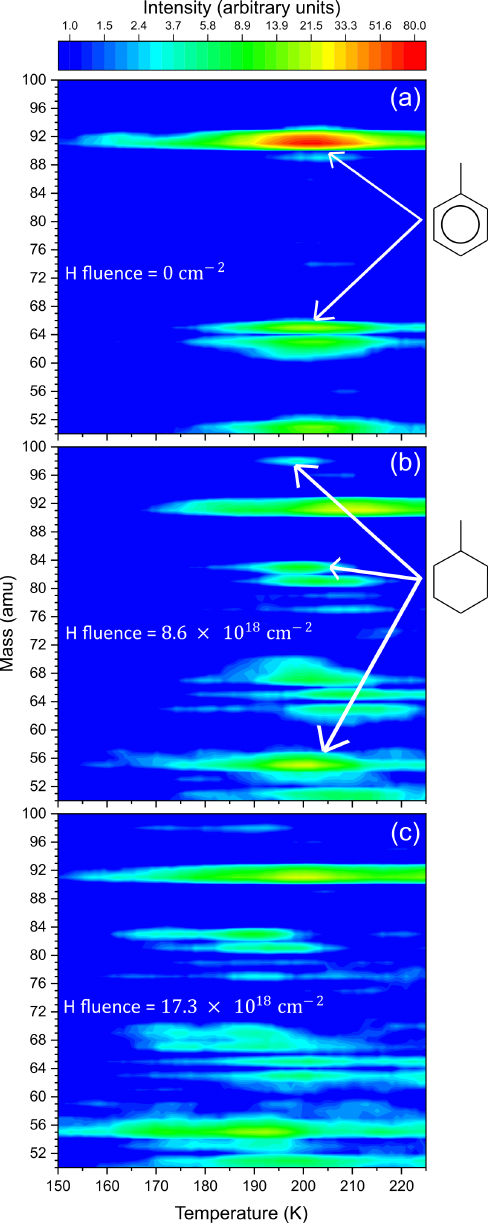}}
    \centering
     \caption{Three contour plots showing the TPD traces for toluene exposed to different fluences of H-atoms. (a) Pristine toluene (92\,amu), and its H-loss fragment (91\,amu) being observed at around 200\,K. (b) TPD of a toluene monolayer exposed to a fluence of $8.6\times10^{18}$\,cm$^{-2}$\,H-atoms showing a reduction in the toluene related desorption features along with the appearance of new desorption products with masses 98 and 83\,amu at lower temperatures, attributed to the formation of methyl-cyclohexane. (c) TPD corresponding to a H-atom fluence of $17.3\times10^{18}$\,cm$^{-2}$ where the hydrogenated species and the toluene now all desorb at lower temperatures.}
     \label{contour}
 \end{figure}

Fig. \ref{contour} shows three contour plots for TPD measurements following the exposure of a monolayer of toluene on HOPG to different fluences of H-atoms. The plots show the intensity of the species in the considered mass range that desorbed from the surface as the surface temperature was increased. Fig.\ref{contour}(a) corresponds to the TPD of pristine toluene \ce{C6H5CH3} and shows several mass detections, with the most intense desorption signal at 200\,K, consistent with the monolayer desorption temperature. The signal at 92\,amu corresponds to the parent ion of toluene, while that at 91\,amu is consistent with the loss of a H-atom during electron impact ionization. This, along with the less intense signals at masses 65, 63 and 51\,amu is consistent with the expected fragmentation pattern for electron impact ionization \citep{TolueneNIST}.

Toluene has six carbon atoms that are not fully hydrogenated and so, similarly to PAHs, it is expected that, when exposed to H-atoms, these carbon atoms can be further hydrogenated. There are stable numbers of additional H-atoms that can be added to toluene that correspond to the addition of pairs of H-atoms. This arises as the addition of a H-atom to the carbon ring leads to an open shell radical system that is highly reactive towards further hydrogen addition. This then leads to a more stable closed-shell system. In Fig.\ref{contour}(b) the TPD measurement performed after a H-atom exposure of $8.6\times10^{18}$\,cm$^{-2}$ is shown. The signatures of pristine toluene at 92 and 91\,amu are still present, but reduced in intensity, indicating some of the initial toluene has reacted. The observation of a species with mass 98\,amu is consistent with the formation of, methyl-cyclohexane \ce{C7H14}, including its expected electron impact ionization fragments at 83, 70, 69, 56 and 55\,amu \citep{methylNIST}. The hydrogenated toluene species desorb at lower temperatures than toluene, with methyl-cyclohexane desorbing at 196\,K. At a slightly higher temperature, the less intense signals at 96, 81\,amu are attributed to the parent and fragment ions of the partially hydrogenated toluene. Fig.\ref{contour}(c) shows the contour plot corresponding to the TPD performed following exposure of toluene to a larger H-atom fluence of $17.3\times10^{18}$\,cm$^{-2}$. The methyl-cyclohexane has a desorption peak at around 170\,K in addition to that at 190\,K. There is also a signal at 84\,amu associated with this lower temperature peak which is attributed to the formation of cyclohexane. The lower temperature methyl-cyclohexane desorption therefore likely results from co-desorption with the cyclohexane. We note that the measured desorption temperatures, while dependent on experimental parameters such as the temperature ramp, typically reflect the desorption energy, thus indicating that methyl-cyclohexane and hexane are bound to the HOPG less strongly than toluene.

\begin{figure}[th]
     \resizebox{0.9\hsize}{!}{\includegraphics{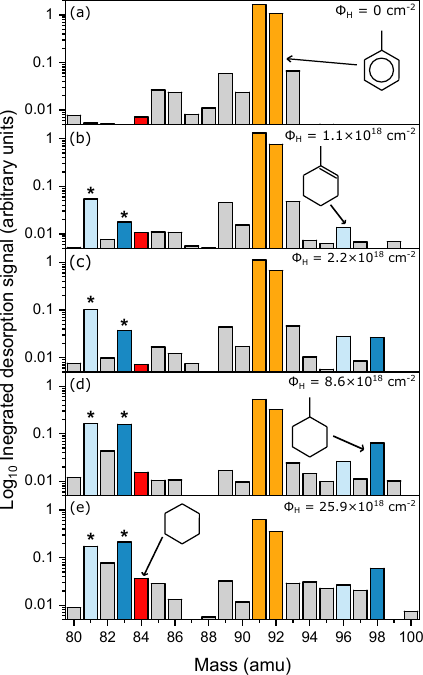}}
     \centering
     \caption{The integrated desorption signals for species with masses 80 to 100\,amu integrated over a desorption temperature range of 150\,-\,215\,K as a function of increasing H-atom fluence. The integrated signals have not been not normalised. Toluene, highlighted in orange decreases while the hydrogenation products, indicated by shades of blue, increase. Fully hydrogenated toluene is observed at 98\,amu. Fragments at -15\,amu relative to the parent ions, indicated by *, are consistent with \ce{CH3} loss during ionization. The peak at 84\,amu, highlighted in red, corresponds to cyclohexane.}
     \label{Bar}
 \end{figure}

Fig.\ref{Bar} shows five mass distributions of desorption products for (a) pristine toluene and (b-e) increasing H-atom exposures up to $\Phi_\text{H}$ = $17.3\times10^{18}\,\text{cm}^{-2}$. The abundance of each mass product was obtained by integrating the relevant TPD trace over a desorption temperature range of 150\,K to 215\,K, in order to exclude desorption signal from the sample mount at higher temperatures, see Appendix Fig.\ref{dep}. This integration range encompasses toluene and all of the formed products, as shown in Fig.\ref{contour}. In Fig.\ref{Bar}(a) the largest integrated desorption signals, shown in orange, are associated with toluene at 92\,amu and the ion corresponding to the loss of one H-atom during electron impact ionization at 91\,amu. In addition to these high mass species, lower mass species (65 and 63\,amu) are also observed, consistent with the expected fragmentation pattern for electron impact ionization of toluene \citep{TolueneNIST}. The parent and fragment ion masses, along with molecular structures and desorption temperatures for toluene and the main hydrogenation products are presented in Table\,\ref{Bp}. 

Toluene contains seven carbon atoms, 6 of which are sp$^2$ hybridized and so, similarly to previously investigated PAHs, it is expected that, when exposed to H-atoms, these carbon atoms will undergo hydrogen addition \citep{rijuthacornaulene}. Fig.\ref{Bar}(b-d) shows that as the H-atom fluence increases the masses 96 and 98\,amu (light and dark blue, respectively) increase in intensity while that for toluene decreases. The 96\,amu mass corresponds to partially hydrogenated toluene where four extra H-atoms have been added, leaving only two unsaturated carbon atoms. The main electron impact ionization fragment ion at 81 amu is also detected (light blue). This fragment is formed in the QMS and corresponds to the removal of the methyl group. The partially hydrogenated species are already observed at hydrogen fluences as low as $\Phi_\text{H}=1.1\times10^{18}\,\text{cm}^{-2}$ in Fig.\ref{Bar}b. Further H-atom exposure results in the formation of fully hydrogenated toluene, i.e. methyl-cyclohexane (\ce{CH3C6H11}), with a mass of 98\,amu (dark blue). This is corroborated by the detection of its main fragment, at a mass of 83\,amu (dark blue), formed via the loss of the methyl group. The other expected fragments from methyl-cyclohexane at 55 and 56\,amu \citep{methylNIST}, are detected as shown in Fig.\ref{contour} and detailed in Table\,\ref{Bp}. The appearance of fragment ions at -15\,amu relative to the parent ion, indicated by * in Fig.\ref{Bar}, for the hydrogenated species is consistent with the loss of the methyl group during electron impact ionization. The detection of mass 84\,amu (red) is ascribed to the formation of cyclohexane on the surface, as discussed in Section \ref{cyclohexform}. Clearly, the abundance of hydrogenated species increases with H-atom fluence. The decrease in the signal corresponding to toluene is not as apparent due to the logarithmic axis. As such, a linearly scaled version can be found in Fig.\ref{Barlin}.

Fig. \ref{Bar} shows that the addition of 2, 4 and 6 H-atoms to toluene leads to particularly abundant species. These species display an increase stability as a result of the odd-even electron rule \citep{magicnumbers} in which a second H-atom adds without a barrier to the radical species formed following the breaking of a \ce{C=C} double bond during the first H-atom addition. This is corroborated by theoretical calculations in Section \ref{theoryresults}, where Fig.\ref{fig:hydrogenation_levels} shows that every second H-atom addition is significantly more thermodynamically favoured than a single H-atom addition.

\begin{table*}[t]
\caption{The three main molecules detected in the TPD measurements with their structure, masses, QMS fragments and desorption temperature.}              
\label{Bp}      
\centering                                      
\begin{tabular}{c c c c c}          
\hline\hline                        
 
Molecule & Structure & Parent mass (amu) & Main QMS fragments (amu) & Desorption temperature (K) \\  
\hline
    \\[-0.8em]
     Toluene \ce{CH3C6H5} & \includegraphics[width=0.5cm,angle=90]{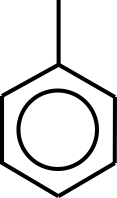}& 92 & 91, 65, 63, 51 & 200 \\ 
     Methyl-cyclohexene \ce{CH3C6H9}&\includegraphics[width=0.5cm,angle=90]{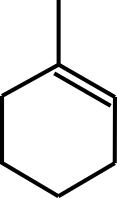}  & 96 & 81, 68, 67, 55 & 196  \\ 
     Methyl-cyclohexane \ce{CH3C6H11}&\includegraphics[width=0.5cm,angle=90]{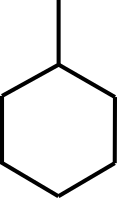}  & 98 & 83, 70, 69, 56, 55 & 190  \\ 
     Cyclohexane \ce{C6H12}& \includegraphics[width=0.5cm,angle=90]{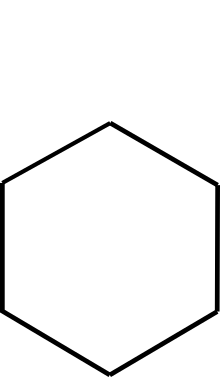}& 84 & 69, 56, 55 & 174  \\
\hline
\end{tabular}
\end{table*}

\subsection{Hydrogen addition cross-section}

\begin{figure}[t]
    \resizebox{0.9\hsize}{!}{\includegraphics{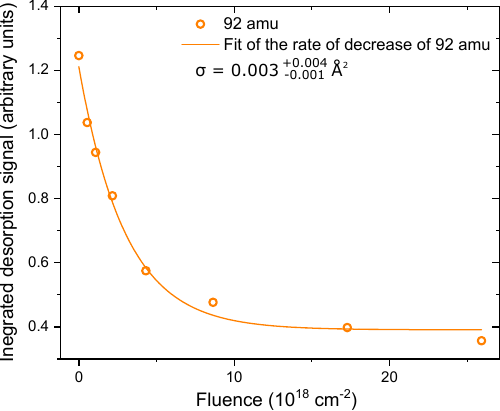}}
    \centering
     \caption{The integrated desorption signal for toluene (92\,amu) as a function of H-atom fluence. The data is fitted with the exponential decay function given in equation \ref{decayform}. The best fit value for the cross-section for the first H-atom addition is $\sigma=0.003^{+0.004}_{-0.001}$\,Å$^2$.}
     \label{rate}
 \end{figure}

Fig.\ref{rate} shows the evolution of the integrated desorption signal of toluene, 92\,amu, when exposed to increasing fluences of H-atoms. This can be fitted with a single exponential decay, consistent with first order reaction kinetics according to 

\begin{equation}\label{decayform}
    I = I_0 \exp(-\Phi_\mathrm{H} \sigma)+I_{\infty},   
\end{equation}

where $I$ is the integrated desorption signal for toluene, $I_0$ is that obtained for a monolayer of toluene in the absence of H-atom exposure, $\Phi_H$ is the H-atom fluence and $I_{\infty}$ accounts for any background toluene signal and possible sample holder contribution. From this fit, the cross-section, $\sigma$, for the addition of the first hydrogen is found to be $\sigma = 0.003  ^{+0.004} _{-0.001}$\,Å$^2$. The derived cross section should be considered a lower limit because of potential cycles of abstraction-addition reactions that may occur throughout the superhydrogenation process. Here hydrogen reacts with an incoming hydrogen to form \ce{H2} and then another incoming H-atom then replaces the one that was lost. This effectively leads to the exchange of one hydrogen for another with no change in mass. This cross section can be compared to that found previously for larger PAHs. For example coronene \ce{C24H12} has a cross-section of  $\sigma = 0.027 ^{+0.027} _{-0.007}$\,Å$^2$ for H-atoms \citep{2019MNRAS.486.5492J}. This shows that toluene has a much smaller H addition cross section and so is hydrogenated much more slowly. This may be due to it consisting of only a single aromatic ring meaning it has a smaller geometric cross-section, with only 6 available H addition sites compared to 24 for coronene, yielding a factor of 4 difference in the number of addition sites. An even larger reduction of the cross section could be expected from consideration of the reduction in the molecular area when going from the 7 carbon ringed coronene molecule to the single carbon ring toluene. In addition the presence of the methyl-group may play a small role in this reduction. We attribute the non-zero toluene signal at infinite H-fluence to the presence of unreacted toluene desorbing from the sample holder. Fig. \ref{dep} clearly shows that the sample holder contribution for the prepared monolayer likely extends into the integration interval. On this basis, we do not expect the derived cross-section for the reaction to be affected by this.

\subsection{Methane and cyclohexane formation}\label{cyclohexform}

\begin{figure}[t]
    \resizebox{0.9\hsize}{!}{\includegraphics{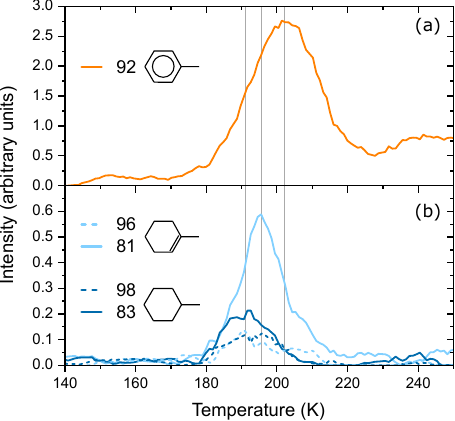}}
    \centering
     \caption{(a) TPD trace for toluene (92\,amu) in orange following a H-atom fluence of $2.2\times10^{18}$\,cm$^{-2}$. (b) TPD traces on a different Y axis scale of partially hydrogenated toluene (parent mass 96\,amu, main fragment 81\,amu) in light blue and methyl-cyclohexane (parent mass 98\,amu, main fragment 83\,amu) in dark blue. The vertical grey lines show the desorption peaks of the three different molecules. The toluene desorbs at the highest temperature at 202\,K, then the partially hydrogenated toluene at 196 K and then the methyl-cyclohexane has the lowest at 191\,K.}
     \label{lowtpd}
 \end{figure}

Fig. \ref{lowtpd} shows individual TPD traces for a toluene monolayer exposed to a H-atom fluence of $2.2\times10^{18}$\,cm$^{-2}$. The integrated desorption signals are those shown in Fig.\ref{Bar}(c). Fig. \ref{lowtpd}(a) shows the trace of the toluene (92\,amu, orange) with a desorption peak at 202\,K. Fig. \ref{lowtpd}(b) shows the traces of its main hydrogenation products with partially hydrogenated toluene (parent 96\,amu, main fragment 81\,amu) in light blue being the main product desorbing at a lower temperature than the toluene at 196\,K. Methyl-cyclohexane i.e. fully hydrogenated toluene (parent mass 98\,amu, main fragment 83\,amu), has been formed and can be seen in dark blue. This species desorbs at a lower temperature of 191\,K showing that increasing the degree of hydrogenation weakens the binding energy of the molecule to the surface leading to a lower desorption temperature as shown in Fig.\ref{contour}. In Fig.\ref{lowtpd} it is observed that all the desorption peaks of the differently hydrogenated toluene species overlap with very little temperature difference between each species. This is likely due to the molecules being similar and having similar binding energies but may also be due to co-desorption where one species desorbing from the surface can cause another to do the same. The main desorption peaks for each species are separated sufficiently in temperature to be identified as separate molecule desorption peaks and the lowering of the desorption temperature with increasing hydrogenation degree can be observed. 

\begin{figure}[t]
    \resizebox{0.9\hsize}{!}{\includegraphics{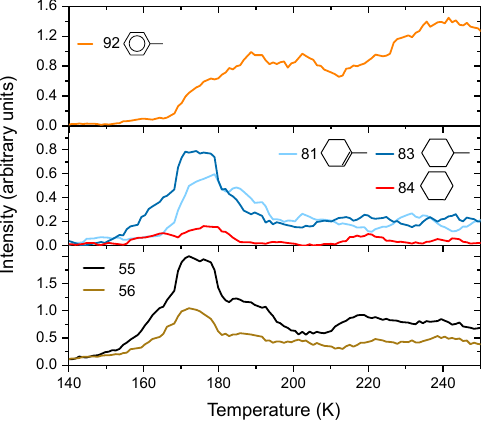}}
    \centering
     \caption{(a) TPD trace for toluene (92\,amu) in orange following a H-atom fluence of $25.9\times10^{18}$\,cm$^{-2}$. (b) TPD traces, on a different y-axis scale, for partially hydrogenated toluene (parent mass 96\,amu, main fragment 81\,amu) in light blue and methyl-cyclohexane (parent mass 98\,amu, main fragment 83\,amu) in dark blue and cyclohexane (84\,amu) in red. The cyclohexane desorbs over a range of temperatures starting from 160\,K. (c) shows the shared QMS fragments of mass 55 and 56\,amu that are fragments of the hydrogenated toluene species and cyclohexane.}
     \label{hightpd}
 \end{figure}

Fig.\ref{hightpd} shows the TPD traces for a toluene monolayer exposed to a higher H-atom fluence of $25.9\times10^{18}$\,cm$^{-2}$. The integrated desorption signals are those shown in Fig.\ref{Bar}(e). Fig.\ref{hightpd}(a) shows that the toluene peak is much less intense, indicating a higher degree of processing through H-atom addition. Fig.\ref{hightpd}(b) shows the traces of the hydrogenation products. The desorption temperature for the partially hydrogenated toluene (light blue) and methyl-cyclohexane (dark blue) is lower than in Fig.\ref{lowtpd}. This is attributed to there being significantly less trapping by toluene than seen in Fig.\ref{lowtpd} because of the reduced surface abundance of toluene. A species with mass 84\,amu, shown in red, starts to desorb at 160\,K, continuing with the other hydrogenation products at 178\,K. The detection of mass 84\,amu is not consistent with any fragment ions for toluene hydrogenation products. The detection of mass 84\,amu at large hydrogen fluences is also seen in Fig.\ref{Bar}. On this basis, we attribute the observation of mass 84\,amu to the formation of cyclohexane (\ce{C6H12}) on the HOPG surface.

Fig.\ref{hightpd}(c) shows the traces of 55 and 56\,amu which are attributed to fragments that are common to both methyl-cyclohexane and cyclohexane. The ratio between these fragments changes for each different molecule \citep{methylNIST,cycloNIST} and it is expected there is more 56\,amu fragmented for cyclohexane and 55\,amu for methyl-cyclohexane. Although it is not possible to directly observe this ratio here. due to co-desorption of the cyclohexane and methyl-cyclohexane, all of the expected fragments of cyclohexane 84 such as 69 and 56\,amu \citep{cycloNIST} are observed.

\begin{figure}[t]
     \resizebox{0.9\hsize}{!}{\includegraphics{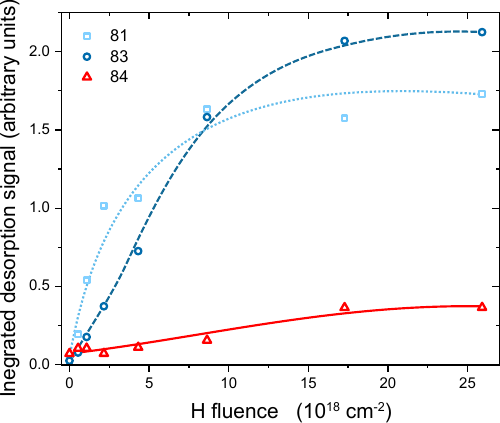}}
     \centering
     \caption{The integrated desorption signals for the main products of toluene hydrogenation as a function of increasing H-atom fluence. Partially hydrogenated toluene (81\,amu) is shown in light blue with. Methyl-cyclohexane (83\,amu) is shown in dark blue. Cyclohexane (84\,amu) is shown in red. The lines are added to guide the eye.}
     \label{prodchange}
 \end{figure}

\begin{figure}[h]
    \resizebox{0.6\hsize}{!}{\includegraphics{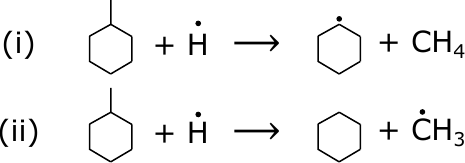}}
    \centering
     \caption{The two possible chemical equations of the reaction that causes the removal of the methyl group. The dot above the molecule or atom indicates a radical. (i) has the formation of methane and a radical cyclohexane. (ii) would be if the cyclohexane added the hydrogen atom and ejected a \ce{CH3} radical. These different routes are investigated in Section \ref{theoryresults}.}
     \label{react}
 \end{figure}

Fig.\ref{prodchange} shows the integrated desorption signal of cyclohexane and the two dominant fragments of the hydrogenated toluene as a function of H-atom fluence. The abundance of species with masses corresponding to partially hydrogenated toluene (81\,amu) increases for low H-atom fluences before saturating for higher H-atom fluences. This confirms they are linked to intermediate superhydrogenation products. The dominant fragment of methyl-cyclohexane (83\,amu) increases more slowly but reaches a larger abundance at the higher fluences before saturating. The signal due to cyclohexane (84\,amu) increases with H-atom fluence more slowly than partially hydrogenated toluene indicating it is formed only at the largest H-atom fluences. This suggests that cyclohexane formation relies on the presence of highly hydrogenated toluene species such as methyl-cyclohexane. In this case an incoming H-atom attacks the \ce{C-C} bond of methyl-cyclohexane between the carbon ring and the methyl group. This could result in fragmentation into \ce{CH4} and a partially H saturated cyclohexane species or a radical \ce{CH3} and a fully H saturated cyclohexane species as shown in Fig.\ref{react}. It should be noted that the sample temperature of 110\,K during H-atom exposure was significantly higher than the typical \ce{CH4} desorption temperature of 55\,K \citep{methane} and thus only the cyclohexane was observed. This is further investigated through theoretical calculations in Section \ref{theoryresults}.

\subsection{Theoretical calculations} \label{theoryresults}

To support the hypothesis that the hydrogenation of toluene can lead to the formation of methyl-cyclohexane, cyclohexane, and methane, theoretical calculations have been conducted using DFT. Firstly, we are interested in quantifying the energy liberated by the incremental hydrogenation of toluene, which is done by sequential addition of hydrogen to the toluene molecule and evaluation of the change in binding energy. Understanding the energy scale involved in these reactions provides context to the feasibility of further "hot" reactions.

\begin{equation}
    E_\mathrm{cohesive} = E_\mathrm{DFT, product} - \sum{E_\mathrm{DFT,reactant_i}}
\end{equation}

\begin{figure}[t]
    \resizebox{0.9\hsize}{!}{\includegraphics{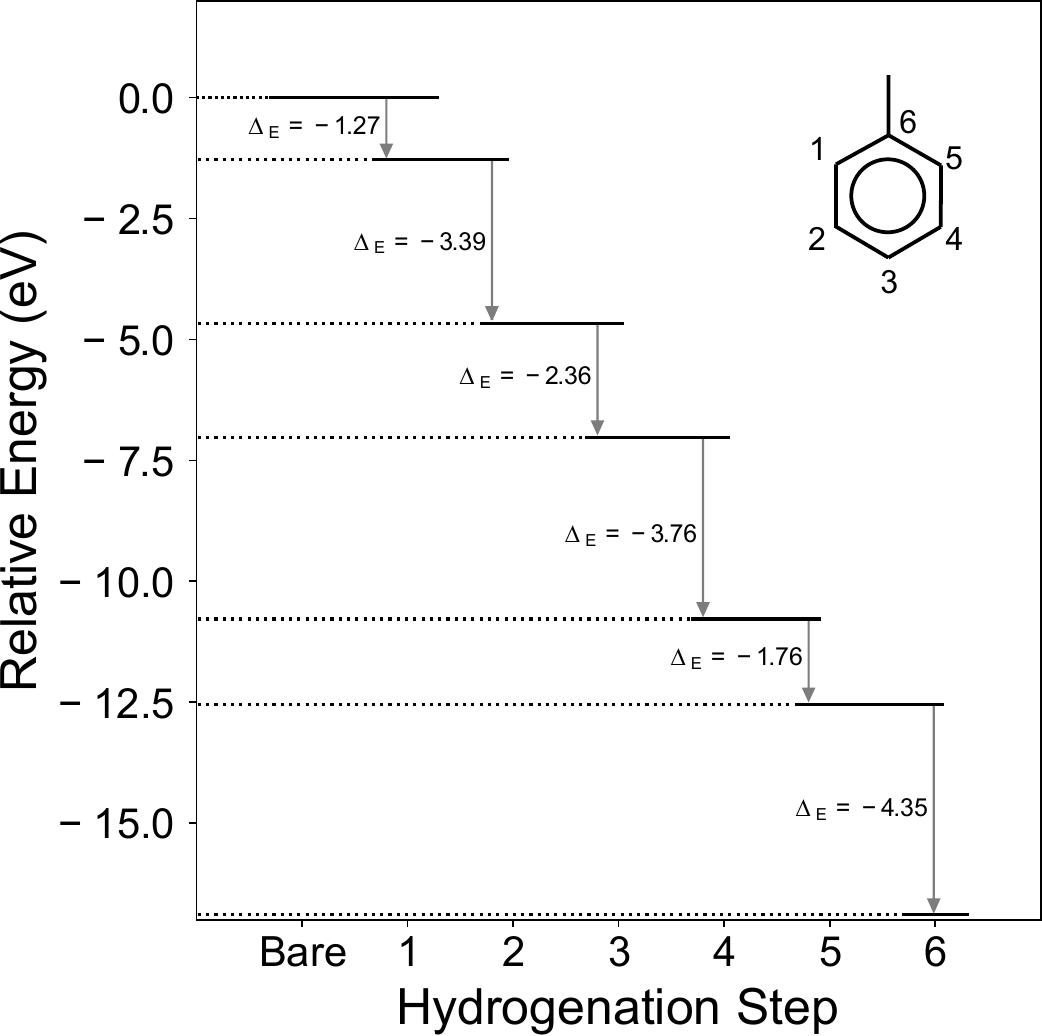}}
    \centering
     \caption{Relative energy levels for the incremental hydrogenation of Toluene as calculated with DFT. The $0$th index, labelled "Bare", highlights the toluene reference, with subsequent indices indicating correspondingly increasing hydrogenation levels. This hydrogenation proceeds from a beta carbon, sequentially around the ring terminating in the alpha carbon, see inset.}
     \label{fig:hydrogenation_levels}
 \end{figure}

The results of this examination are displayed in Fig.\ref{fig:hydrogenation_levels}. Each line indicates the total amount of energy gained, relative to the energy of toluene, for the given hydrogenation step and all preceding steps. Hydrogen was first added to a beta carbon, and then successively around the ring, concluding with the alpha carbon. Of particular interest is the substantial chemical energy liberated between reaction steps 1-2, 3-4 and 5-6, with the most notable being the final 5-6 step. At the conclusion of this final hydrogenation, $>$\,4\,eV of energy is liberated. We find strong agreement with other theoretical works \citep{theyrpaper}. 

\begin{figure}[t]
    \resizebox{0.9\hsize}{!}{\includegraphics{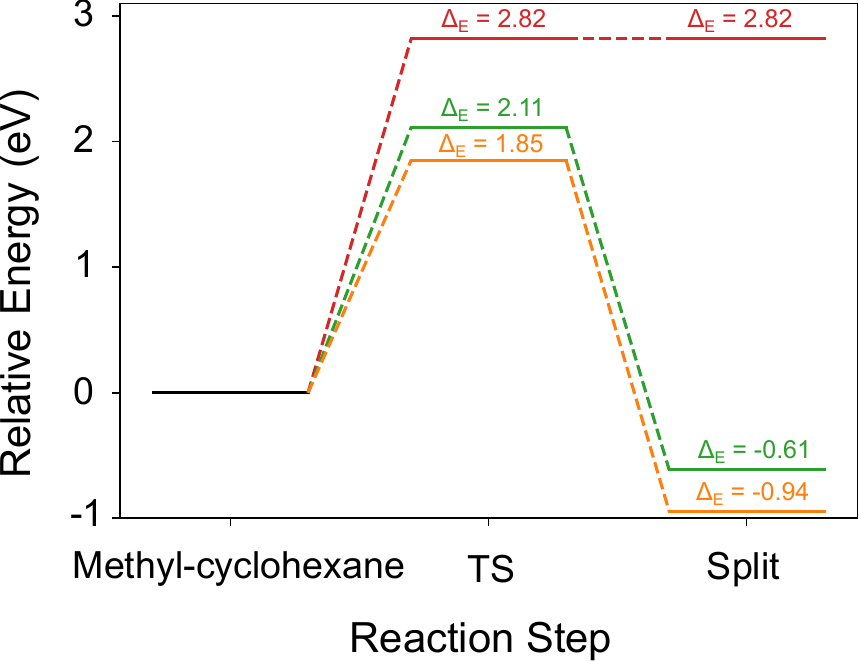}}
    \centering
     \caption{Calculated barriers for spontaneous scission of fully hydrogenated toluene (red) into \ce{^{.}C6H11} and \ce{^{.}CH3}. Scission of fully hydrogenated toluene into \ce{^{.}CH3} and cyclohexane given incidence of hydrogen onto the alpha carbon (green), see Fig. \ref{react}(ii). Scission of fully hydrogenated toluene into \ce{^{.}C6H11} and CH$_4$ given incidence of atomic hydrogen upon the methyl group (orange), see Fig. \ref{react}(i). The energies for these products and the corresponding transition states are labelled relative to methylcyclohexane.}
     \label{fig:scission}
\end{figure}

From a thermodynamic viewpoint, full hydrogenation of the carbon ring of the toluene is energetically favourable. We now consider pathways for some potential reactions occurring following full hydrogenation, by which the methyl group might be liberated in some form. We consider multiple potential mechanisms for such a reaction to occur, and identify 3 plausible reactions from the calculated MEPs. These are displayed in Fig. \ref{fig:scission}. The first, and simplest, of the plausible reactions is an immediate scission of fully hydrogenated toluene into two radical groups, both radical cyclohexane and \ce{^{.}CH3} (shown in red). This reaction is endothermic and not thermodynamically incentivised, although it lies within the bounds of the liberated chemical energy from a single hydrogenation step, most notably that released during the $6^{\mathrm{th}}$ hydrogenation step. This indicates that if the liberated energy is not quickly delocalized, such processes could lead to the release of two radical species. 

Continual exposure to H-atoms opens for the possibility of further reactions, namely scission via the formation of a further C-H bond at the cost of the \ce{C-C} at the adjoining carbon. Both considered reactions, (\ce{CH4} formation and \ce{^{.}CH3} formation, with radical and non-radical cyclohexane respectively, see Fig. \ref{react}) are exothermic. However, the formation of \ce{CH4} (shown in orange) is more so. Additionally, the barrier for this scission is lower in the case of \ce{CH4} formation, indicating a preference for methane formation over its radical counterpart. Furthermore, sterically the liberation of \ce{^{.}CH3} (shown in green) requires that hydrogen be incident upon the methyl-bonded carbon from the correct side (i.e. not such that the hydrogen bonded to said carbon be impacted by the incidence). Thus, we speculate that this reaction becomes even more challenging when the orientation of the molecule is considered relative to a surface. No such requirement is observed for the methyl group, which is geometrically more exposed to reaction.

The experimental observation of cyclohexane formation, combined with the theoretical calculations shows that the final hydrogenation step liberates sufficient energy to overcome the barrier of breaking the \ce{C-C} bond. With another atom present such as a hydrogen this can result in the formation of a methane and a radical carbon ring which can then react with a further H-atom to form cyclohexane.

\section{Astrochemical implications and conclusion}

We have shown that toluene can interact with H-atoms, leading to fully superhydrogenated toluene as has been observed for larger PAH molecules. The cross section for the first addition of hydrogen to toluene is found to be $\sigma = 0.003  ^{+0.004} _{-0.001}$\,Å$^2$. This is compared to coronene which has a cross-section of H addition of  $\sigma = 0.027  ^{+0.027} _{-0.007}$\,Å$^2$ \citep{2019MNRAS.486.5492J}, which is nine times larger. This is almost consistent with the difference in geometric cross-section between the two molecules, where coronene consists of seven carbon rings and toluene only one. Thus, the presence of the methyl group does not significantly change the initial intrinsic reactivity towards hydrogen atoms, with H-atom addition to the carbon ring proceeding akin to to the outer edge sites of coronene. This suggests that the general process of hydrogen addition, as well as \ce{H2} forming abstraction reactions, proceeds in a similar way for methyl- as for the non-methylated PAHs. The degree of hydrogenation achieved in the ISM will depend on the balance between H-addition and UV induced dehydrogenation. It is known that UV photons of sufficient energies can cause the removal of H-atoms or molecular hydrogen \citep{PAHsandUV} and therefore this is expected to occur similarly for toluene. Due to this it is expected that, in regions with UV radiation, only low degrees of superhydrogenation will be reached and it will act similarly to PAHs catalysing molecular hydrogen formation. In regions shielded from the UV photons however it is expected that high degrees of superhydrogenation will be reached.

Our measurements demonstrate that, for high H-atom fluences, the \ce{CH3} group can be detached from toluene, with H-atom addition to the cyclohexane radical leading to the observed cyclohexane formation. Our calculations indicate that the the scission of the \ce{C-CH3} bond might be facilitated by the liberation of around 4.2\,eV through hydrogenation of the final site on the aromatic ring. This is sufficient energy to allow for several "hot" reactions to occur, provided that this energy is not dissipated prior to further reaction. Firstly, the toluene can spontaneously split into two radicals without the addition of further hydrogen, an endothermic process. Alternatively, further H-atom addition can also lead to scission, either leading to \ce{CH4} and \ce{^{.}C6H11}, or \ce{C6H12} and \ce{^{.}CH3}, with the former being both thermodynamically and kinetically preferred. We note that scission into \ce{^{.}CH3} is less favourable due to steric considerations, something we expect to be exacerbated by the presence of a surface.

It has been suggested that toluene and small methylated PAH species play a role as intermediates in the MAC and MACA formation mechanisms leading to larger PAHs \citep{Shukla2010, Doddipatla2021}. The survival of the methyl group is therefore an important factor for PAH growth. Our measurements show that demethylation likely occurs after significant hydrogenation of the methyl carrying ring, and is therefore likely to be more significant for smaller methyl PAHs. The intrinsic hydrogenation efficiency does not appear to be strongly affected by the presence of the methyl group, although further studies would be required to consider how this extends to larger, more highly methylated PAHs. Thus, the intricate balance between H-atom and UV fluxes will determine the extent to which PAH growth occurs, dictating the size of the PAH molecules formed.

The observed hydrogenation induced loss of the methyl group is, in itself, of interest for interstellar chemistry involving methylated PAHs. The removal of \ce{^{.}CH3} could provide a route to \ce{CH4} formation and leaves a radical carbon ring. On a grain surface, such radical species could react with other molecular species providing a route to form ring containing complex organic molecules. For example, cyclohexane is susceptible to \ce{C-C} band cleavage by nitrogen and oxygen \citep{GEKHMAN2004833}, potentially leading to the formation of complex organic molecules of prebiotic interest such as alcohols and fatty acids and nitrogen bases. 

To summarise, our results demonstrate that, when exposed to H-atoms, toluene can become fully hydrogenated. The cross-section for the first H-atom addition is nine times smaller than that of coronene, largely consistent with the difference in geometric cross-section. This implies that H-addition reactions to the carbon ring are not strongly affected by the presence of the methyl group. Toluene can become fully hydrogenated, resulting in methyl-cyclohexane. The exothermicity of the final hydrogenation step may be sufficient to lead to scission of the \ce{C-CH3} bond which, in the presence of an additional H-atom, most favourably leads to the formation of \ce{CH4} and \ce{^{.}C6H11}. The \ce{^{.}C6H11} radical could then react with a further H-atom to form cyclohexane, as observed in the experiments, or with other ice species, potentially leading to the formation of complex organic molecules.

\begin{acknowledgements}
We acknowledge support by the Danish National Research Foundation through the Center of Excellence “InterCat” (Grant Agreement No. DNRF150) and by the VILLUM FONDEN through Investigator Grant, Project No. 16562.
\end{acknowledgements}

\bibliographystyle{aa} 
\bibliography{bibliography}

\begin{appendix}

\section{Preparation of the toluene monolayer}\label{tol_mono}
\begin{figure}[h]
     \resizebox{0.9\hsize}{!}{\includegraphics{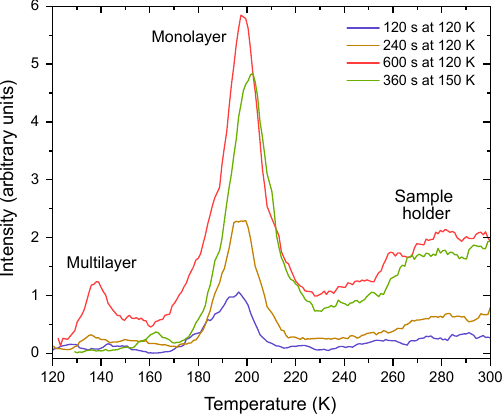}}
     \centering
     \caption{TPD traces showing four depositions of toluene on HOPG. The signals correspond to the parent ion of toluene ($m/z=92$). The first three traces correspond to toluene deposited on HOPG at 120\,K for 120\,s (blue), 240\,s (dark orange), and 600\,s (red). The green trace shows a 360\,s toluene deposition with the HOPG held at 150\,K. With increasing deposition time the monolayer peak around 200\,K grows, followed by the multilayer peak at 138\,K for longer deposition times. The broad peak that appears above 240\,K is attributed to desorption of toluene from the sample holder. When toluene is deposited at 150\,K only the monolayer peak is present and shifted to higher temperatures due to reordering.}
     \label{dep}
 \end{figure}

TPD measurements were performed in order to determine a procedure by which to prepare a single monolayer of toluene on HOPG. Fig. \ref{dep} shows the TPD traces for several different deposition times of toluene on HOPG. The signals correspond to the QMS detection of the parent ion of toluene ($m/z$\,=\,92). Three depositions were performed, with the HOPG at 120\,K, to illustrate how the toluene coverage increases with deposition time (blue, dark orange and red). This is initially observed through an increasing intensity of the monolayer desorption peak at around 200\,K. For the longest toluene deposition, of 600\,s (red), the formation of multilayers is observed through the appearance of a peak at a lower temperature of ca. 138\,K. The toluene signal at higher temperatures (>\,240\,K) is attributed to desorption from the sample holder. Based on these TPD measurements, a monolayer was prepared by depositing toluene with the HOPG held at a temperature of 150\,K, i.e., above the multilayer desorption. The TPD trace for the resulting saturated monolayer is shown in green in Fig. \ref{dep}. The peak of the monolayer deposited at this 150\,K is shifted to a slightly higher temperature. We attribute this to a reordering of the toluene molecules on the surface, which increases their binding energy.

\section{Integrated desorption signals}

\begin{figure}[th]
     \resizebox{0.9\hsize}{!}{\includegraphics{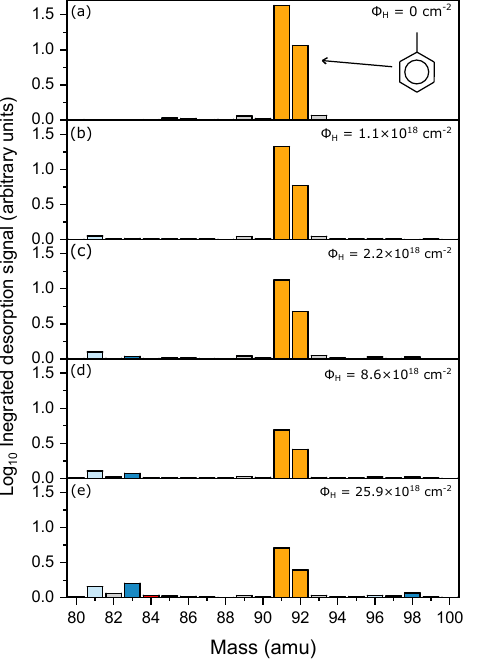}}
     \centering
     \caption{The integrated desorption signals for species with masses 80 to 100\,amu integrated over a desorption temperature range of 150\,-\,215\,K as a function of increasing H-atom fluence. Toluene, highlighted in orange, decreases while the hydrogenation products, indicated by shades of blue, increase.}
     \label{Barlin}
 \end{figure}

Fig. \ref{Bar} in the main text shows the integrated desorption signals of masses 80 to 100\,amu on a logarithmic scale. This enables the integrated signals resulting from the reaction products to be more easily observed. Fig. \ref{Barlin} shows the same figure but with the integrated desorption signals on a linear scale. This shows more clearly that, with increasing H-atom exposure, the mass signals associated with toluene, i.e., 91 and 92\,amu, decrease. This is due to toluene becoming hydrogenated and forming other molecules. This decrease in the integrated signal of toluene is depicted in Fig. \ref{rate}. However, even for the largest H-atom exposures, as shown in Fig. \ref{Barlin}(e), there is still some toluene present. This could be due to the presence of unreacted molecules on the surface, but is likely mainly due to the desorption of toluene from the sample holder. We cannot exclude that the desorption peak corresponding to this, highlighted in Fig. \ref{dep}, extends into the integration range of 150\,-\,215\,K.

\end{appendix}
\end{document}